\documentstyle[12pt,psfig,a4]{article}
\begin{document}
\markboth{J. V. CUNHA, L. MARASSI and R. C. SANTOS}{}
%
%
\title{NEW CONSTRAINTS ON THE VARIABLE EQUATION OF STATE PARAMETER FROM X-RAY GAS MASS FRACTIONS AND SNE Ia}
\author{J. V. CUNHA$^\ast$, L. MARASSI$^\dag$ and R. C. SANTOS$^\ddag$}

\maketitle

\begin{abstract}
Recent measurements are suggesting that we live in a flat Universe
and that its present accelerating stage is driven by a dark energy
component whose equation of state may evolve in time. Assuming two
different parameterizations for the function $\omega(z)$, we
constrain their free parameters from a joint analysis involving
measurements from X-Ray luminosity of galaxy clusters and SNe type
Ia data.
\end{abstract}

\footnote {Departamento de F\'{\i}sica, Universidade Federal do
Rio Grande do
Norte, CEP 59072-970, Natal, Rio Grande do Norte, Brazil\\
      $^\ast$ jvital@dfte.ufrn.br \\ $^\dag$ luciomarassi@yahoo.com \\
      $^\ddag$ rose@dfte.ufrn.br}

\section{Introduction}    

 \hspace{0.5cm}In the framework of general relativity, the present accelerating
 stage of the Universe (as indicated by
 SNe type Ia observations) can be
 explained by assuming the existence of a substantial amount of an
 exotic dark energy component with negative pressure, also known as
 quintessence\cite{Lima04,Ries04}. The existence of this extra component
 filling the Universe (beyond the cold dark matter) has also been indirectly
suggested by independent studies based on fluctuations of the 3K
relic radiation, large scale structure, age estimates of globular
clusters or old high redshift objects, as well as by the X-ray
data from galaxy clusters\cite{data,Allen02}.

A cosmological constant ($\Lambda$), the oldest and by far the
most natural candidate for dark energy, faces some theoretical
difficulties. The most puzzling of them is the so-called
cosmological constant problem: the present cosmological upper
bound, $\Lambda_o/8\pi G \sim 10^{-47} GeV^{4}$, differs from
natural theoretical expectations from quantum field theory, $\sim
10^{71} GeV^{4}$, by more than 100 orders of magnitude. Actually,
such a problem has also inspired many scenarios driven by a
$\Lambda(t)$ or a time varying decaying vacuum with constant
equation of state\cite{LT96}. Among the remaining candidates to
dark energy, the most promising ones lead to a time dependent
equation of state (EOS), usually associated to a dynamical scalar
field component. Such quintessence models may also parametrically
be represented by an equation of state, $\omega(z)$, as proposed
by Cooray and Huterer\cite{CoHu99}, as well as the one discussed
by Linder\cite{Lin03}, and, independently, by Padmanabhan and
Choudhury\cite{PaCh03}. In principle, the time variation of the
EOS parameter, $\omega(z) \equiv p/\rho$, may allow a clear
distinction between a cosmological constant model and the one
driven by a rolling scalar field.

In actual fact, the exploration of the expansion history of the
universe using $\omega(z)$ gave origin to a heated debate with
growing interest in the recent literature. The SNe type Ia test is
the most promising one related to this subject. However, Maor et
al.\cite{MaBrSt01}, and Weller and Albrecht\cite{WeAl01}, have
also observed  that in order to constrain the evolution of the EOS
with SNe observations, it is necessary to use a tight prior on the
mean matter density of the Universe. A natural way to circumvent
such a problem is to consider  the constraints on the density
parameter from measurements of the X-Ray luminosity of galaxy
clusters together in a joint analysis involving SNe Ia
observations.

In this work we investigate the cosmological implications from
X-ray of galaxy clusters and SNe data by considering two different
classes of EOS evolving with redshift. In the first scenario
(hereafter Model 1), the EOS parameter is defined by\cite{CoHu99}
\begin{equation}
Model\,\, 1 : \,\,\, \omega(z) = \omega_o + \omega_1 z,
\end{equation}
whereas in the second, the EOS parameter reads\cite{Lin03,PaCh03}
\begin{equation}
Model\,\, 2: \,\,\, \omega(z) = \omega_o + \frac{\omega_1 z}{1 + z}
\end{equation}
where $\omega_o$ and $\omega_1$ are constants.

It should be noticed that the linear expression of model 1 yields
a reasonable approximation for most quintessence models out to
redshift of a few, and, of course, it should be exact for models
where $\omega(z)$ is a constant or varies slowly. The unsuitable
aspect of the first expression is that it  grows with no limit at
high  redshifts $z>1$ (for example, distance to the last
scattering surface at $z_{lss}\simeq 1100$). In order to fix such
a problem, some authors\cite{Lin03,PaCh03} have proposed the
second form which has the advantage of giving finite $\omega(z)$
for all $z$. In both cases, $\omega_o$ is the current value of the
EOS parameter and $\omega_1$ defines its variation rate for $R$
close to the present epoch ($z=0$).

The paper is outlined as follows. Next section we set up the FRW
equations for both parameterizations and the angular diameter
distance necessary for our analysis using the X-ray luminosity of
galaxy clusters. In section 3 we obtain our main results, namely:
the observational constraints to the free parameters trough a joint
analysis involving X-ray luminosity of galaxy clusters and SNe type
Ia data. Finally, in section 4 we summarize the main conclusions and
perspectives.

\section{Basic Equations}

\hspace{0.5cm}In what follows it will be assumed that the Universe
is flat and its dynamics  is driven by a pressureless cold dark
matter (CDM) fluid plus a quintessence component. Both components
are separately conserved and the EOS parameter of the quintessence
component is represented by one of the parameterizations appearing
in the introduction (see Eqs. (1) and (2)). By integrating the
energy conservation laws for each component and combining the result
with the FRW equation,  it is straightforward to show that the
Hubble parameter for both models can be written as:
\begin{equation}
\label{mod1} H^{2}_{Model 1} = H_{o}^{2}\left[\Omega_{\rm{M}}(1
+z)^{3} + (1 - \Omega_{\rm{M}})(1 + z)^{3(1 + \omega_0 -
\omega_1)}e^{3\omega_1 z}\right],
\end{equation}
and
\begin{equation}
\label{mod2} H^{2}_{Model 2} = H_{o}^{2}\left[\Omega_{\rm{M}}(1 +
z)^{3} + (1 - \Omega_{\rm{M}})(1 + z)^{3(1 + \omega_0 +
\omega_1)}e^{-3\omega_1(z/1+z)}\right],
\end{equation}
where the subscript $``o"$ denotes a present day quantity and
$\Omega_M$ is the CDM density parameter.

On the other hand, the first attempts involving gas mass fraction
as a cosmological test were originally performed by
Pen\cite{Pen97} and Sasaki\cite{Sas96}, and further fully
developed by Allen {\it et al.}\cite{Allen02} who analyzed the
X-ray observations for six relaxed lensing clusters observed with
Chandra in the redshift interval $0.1 < z < 0.5$. A similar
analysis has also been done for conventional quintessence models
having constant EOS parameter by Lima {\it at al.}\cite{LiCuAl03}.
These authors also discussed the case for a cosmological scenario
driven by phantom energy ($\omega < -1$). Further, this test was
also applied in the context of a Chaplygin gas EOS\cite{CuAlLi04}.
More recently, a detailed analysis using an improved sample
observed with Chandra (26 clusters) was performed by Allen and
collaborators\cite{Alen04} also considering a constant EOS
parameter. In such studies, it is usually assumed that the
baryonic gas mass fraction in galaxy clusters provides a fair
sample of the distribution of baryons in the universe. In what
follows, the gas mass fraction is defined as\cite{LiCuAl03,Alen04}
\begin{eqnarray}
\label{fgass} f_{\rm gas}(z_{\rm i}) = \frac{b \Omega_{\rm b}}
{\left(1+0.19 h^{3/2}\right) \Omega_{\rm M}} \left[ 2h \,
\frac{D_{\rm A}^{\rm{SCDM}}(z_{\rm i})}{D_{\rm A}^{\rm{DE}}(z_{\rm
i})} \right]^{1.5},
\end{eqnarray}
where $b$ is a bias factor motivated by gas dynamical simulations
that takes into account the fact that the baryon fraction in
clusters seems to be lower than for the universe as a whole,
$\Omega_{\rm b}$ stands for the baryonic mass density parameter,
with the term $(2h)^{3/2}$ representing the change in the Hubble
parameter between the default cosmology and quintessence scenarios
while the ratio ${D_{\rm A}^{\rm{SCDM}}(z_{\rm i})}/{D_{\rm
A}^{\rm{DE}}(z_{\rm i})}$ accounts for deviations in the geometry of
the universe from the Einstein-de Sitter CDM model.

In order to derive the constraints from X-ray gas mass fraction in
the next section we shall use the concept of angular diameter
distance, $D_A(z)$. Such a quantity is readily derived in the
present context (see, for instance, Refs. [14] and [15]):
\begin{equation}
\label{Diam} D_{\rm A}^{\rm{DE}} = \frac{H_o^{-1}}{(1 +
z)}\int_{x'}^{1} {dx \over x^2 H(x)}.
\end{equation}
where $x = {R(t) \over R_o} = (1 + z)^{-1}$ is a convenient
integration variable. In what follows, we will consider in our
statistical modelling only flat cosmological models with Gaussian
priors of $h=0.72\pm0.08$\cite{F01} with
$b=0.824\pm0.089$\cite{Alen04} and $\Omega_{\rm
b}h^2=0.0214\pm0.002$\cite{KIRK03}.

\section{Observational Constraints}
\hspace{0.5cm}Let us now discuss the constraints from X-ray
luminosity of galaxy clusters and SNe type Ia data. First, it is
worth notice that the $26$ clusters cataloged by Allen {\it et
al.}\cite{Alen04} are all regular, relatively relaxed systems for
which independent confirmation of the matter density parameter
results is available from gravitational lensing studies.

In order to determine the cosmological parameters $\omega_o$ and
$\omega_1$ we use a $\chi^{2}$ minimization for the range of
$\omega_0$ and $\omega_1$ spanning the interval [-2.3,-0.4] and
[-4,6.5], respectively, in steps of $0.025$. The $68.3\%$, $90.0\%$
and $95.4\%$ confidence levels are defined by the conventional
two-parameters $\chi^{2}$ levels $2.30$, $4.61$ and $6.17$,
respectively. It is very important to note that we do not consider
any prior in $\Omega_{\rm M}$, as usually required by the SNe Ia
test.

In addition to our gas mass fraction analysis we consider the SNe
Ia measurements as given by Riess {\it et al.}\cite{Ries04}. The
best fit of the model of Eq.(\ref{mod1}) is $\chi^2 = 202.06$,
$\omega_o=-1.25$, $\omega_1=1.3$ and $\Omega_{\rm M}=0.26$. For
the Model 2,  the best fit is $\chi^2 = 202.02$, $\omega_o=-1.4$,
$\omega_1=2.57$ and $\Omega_{\rm M}=0.26$.

We now present our joint analysis (X-Ray luminosity from galaxy
clusters and SNe Ia data). In the first EOS parameter we find at
$2\sigma$ of likelihood that $-1.78\leq \omega_o\leq -0.82$ and
$-1.2\leq \omega_1\leq 2.7$. For the other model we get $-2.0\leq
\omega_o\leq -0.8$ and $-2.0\leq \omega_1\leq 5.5$ with $2\sigma$.
In Fig. 1, we show contours of constant likelihood in the
$\omega_o$-$\omega_1$ plane. Note that the allowed range for both
$\omega_o$ and $\omega_1$ is reasonably large thereby showing the
impossibility of placing restrictive limits on these quintessence
scenarios. However, these limits are better than those obtained by
a simple SNe Ia analysis since in this case, the uncertainties on
both parameters are strongly correlated when one marginalizes over
$\Omega_M$.

At this point, it is interesting to compare our results with other
recent determinations of $\omega_o$ and $\omega_1$ derived from
independent methods. For example, the age constraints recently
derived by Jain and Dev\cite{JaDe05} are $\omega_o\leq-0.31$ and
$\omega_1\leq0.96$ for the first model, and $\omega_0\leq-0.31$
and $\omega_1\leq3.29$ for the second one. Riess {\it et
al.}\cite{Ries04} found $\omega _{o}=-1.31_{-0.28}^{+0.22}$
($1\sigma$) and $\omega _{1}=1.48_{-0.90}^{+0.81}$ ($1\sigma$)
with the uncertainties in both parameters strongly correlated. In
the article of Padmanabhan and Choudhury we must to use a constant
$\Omega _{M}$ in order to analyze the $\omega _{o}-\omega _{1}$
plane. It should be stressed  that the EOS corresponding to the
cosmological constant is within the $1\sigma $ contour for
$0.21<\Omega _{M}<0.41$, and models with $\omega _{o}>-1/3$ are
ruled out at a high significance level for $\Omega _{M}<0.4$ (we
must to have very high negative values of $\omega _{1}$ in this
case, and despite the high uncertainties in $\omega _{1}$ present
in this data set, we know that it cannot vary but a few from
$\omega _{o}$); this supernova observation analysis clearly
indicates that the data are not sensitive to $\omega_{1}$ as
compared to $\omega_{o}$.

\begin{figure}[t]
\centerline{\psfig{figure=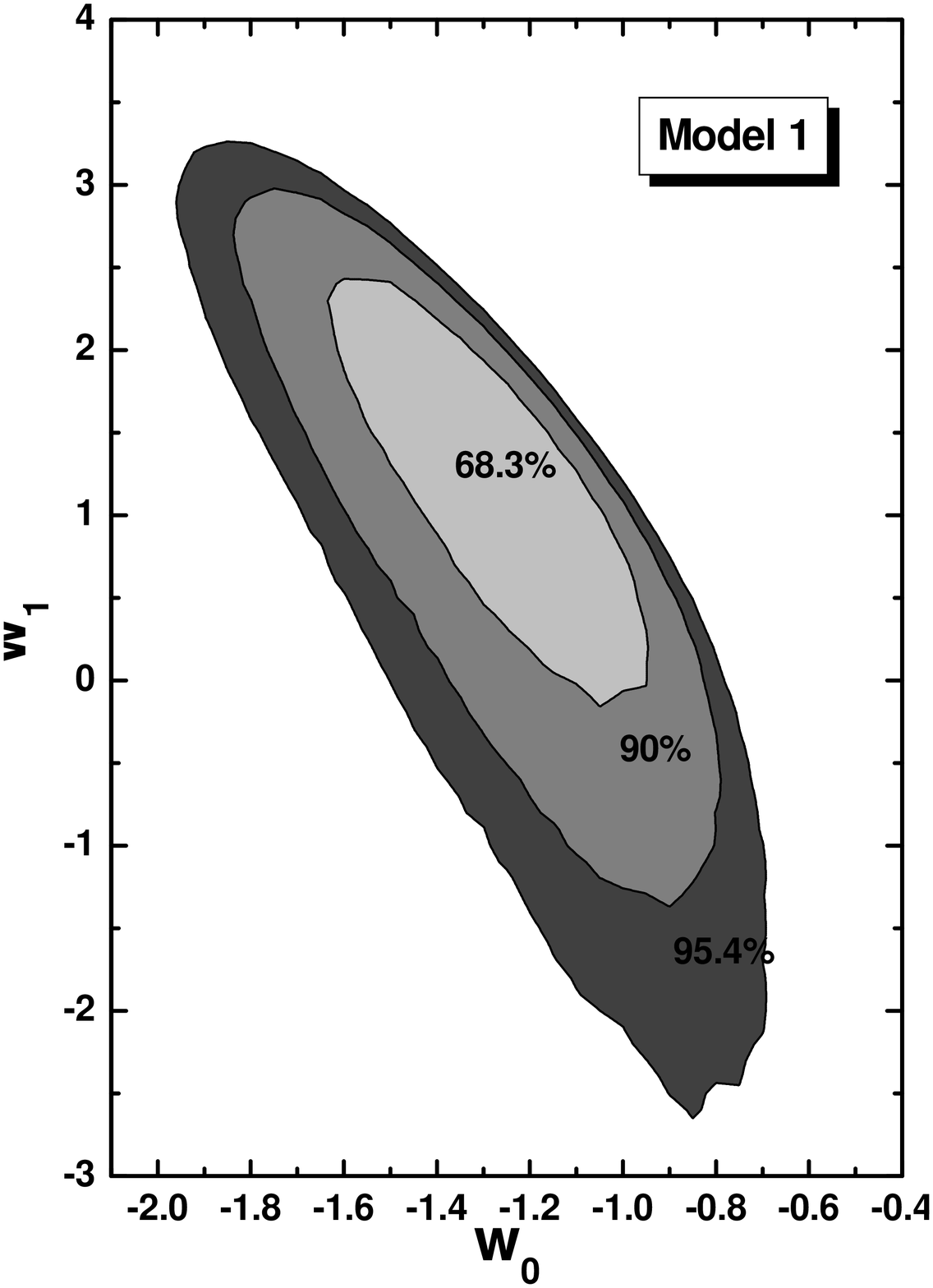,width=2.2truein,height=2.0truein}
\psfig{figure=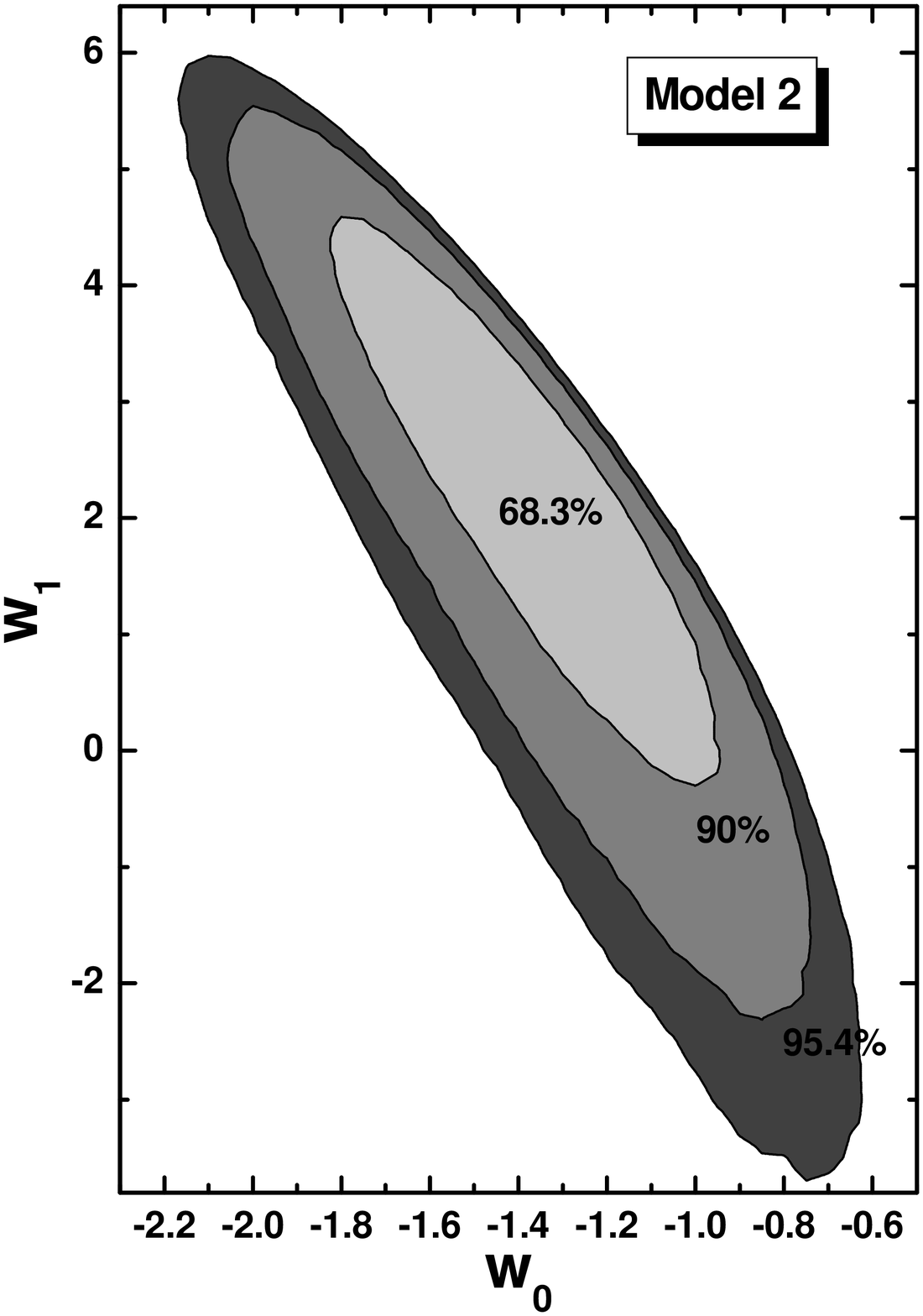,width=2.2truein,height=2.0truein}}
\caption{Marginalized constraints on plane $\omega_0$ and
$\omega_1$ from joint analysis of the Chandra $f_{\rm gas}(z)$ and
SNe Ia data shown above for models 1 (left panel) and 2 (right
panel). The solid lines mark the $1$, $2$ and $3$ $\sigma$
confidence limits. See text for more details.}\label{fig:redshift}
\end{figure}

\section{Conclusion and Perspectives}

Nowadays, the signature of a dark energy or quintessence component
is the most impressive observational fact of cosmology. As
remarked elsewhere, we are living at a special moment where the
emergence of new kind of ``standard cosmology" seems to be
inevitable. In the last few years, a growing attention has been
paid for models with a time varying EOS parameter $\omega(z)$.
With basis on this sort of cosmological scenario, we have
discussed here two simple possible parameterizations of the EOS
obeyed by quintessence models as recently presented in the
literature. Our results suggest that it is worthwhile to use the
estimates of the gas mass fraction from galaxy clusters in joint
analysis with the SNe Ia data since the derived constraints for
$\Omega_M$ (and other quantities) do not require any prior to this
parameter. More important, we have also obtained constraints for
$w_o$ and $w_1$ which have not been obtained before without a
prior in $\Omega_M$.

The parameterizations seems to be more efficient to explain these
data set, once they get a lower $\chi^2$; however they also have an
additional parameter, and that is worthy of a study with more
specific statistical criteria (Akaike or Bayesian information
criteria, for example). A more detailed analysis of this kind will
be investigated in the near future.

\section*{Acknowledgements}
The authors are grateful to Prof. J. A. S. Lima for helpful
discussions. This work was supported by CAPES (Brazilian Research
Agency).

\end{document}